\documentclass[letter]{aa} 
\usepackage{graphicx}
\usepackage{txfonts}
\usepackage{booktabs}
\begin{document} 

   \title{The inhomogeneous sub-millimeter atmosphere of Betelgeuse}

   \author{E. O'Gorman
          \inst{1}\fnmsep\thanks{ogorman@cp.dias.ie},
          P. Kervella\inst{2,3}
          \and
          G. M. Harper\inst{4}
          \and
          A. M. S. Richards\inst{5}
          \and
          L. Decin\inst{6}
          \and 
          M. Montargès \inst{7}
          \and
          I. McDonald\inst{5}
          }
   \institute{Dublin Institute for Advanced Studies, 31 Fitzwilliam Place, Dublin 2, Ireland
              \and
   			   Unidad Mixta Internacional Franco-Chilena de Astronom\'ia (CNRS UMI 3386), Departamento de Astronom\'ia, Universidad de Chile, Camino El Observatorio 1515, Las Condes, Santiago, Chile
              \and
              LESIA (UMR 8109), Observatoire de Paris, PSL Research University, CNRS, UPMC, Univ. Paris-Diderot, 5 place Jules Janssen, 92195 Meudon, France
              \and
              Center for Astrophysics and Space Astronomy, University of Colorado, 389 UCB, Boulder, CO 80309, USA
              \and
              JBCA, Department Physics and Astronomy, University of Manchester, Manchester M13 9PL, UK
              \and
              Institute of Astronomy, KU Leuven, Celestijnenlaan 200D B2401, 3001 Leuven, Belgium
              \and
              Institut de Radioastronomie Millimétrique, 300 rue de la Piscine, 38406 Saint-Martin d'Hères, France
              }

   \date{Received May 15, 2017; accepted June 7, 2017}

  \abstract
   {The mechanisms responsible for heating the extended atmospheres of early-M spectral-type supergiants are poorly understood. So too is the subsequent role these mechanisms play in driving the large mass-loss rates of these stars. Here we present ALMA long (i.e., $\sim$16\,km) baseline 338\,GHz (0.89\,mm) continuum observations of the free-free emission in the extended atmosphere of the M2 spectral-type supergiant Betelgeuse. The spatial resolution of 14\,mas exquisitely resolves the atmosphere, revealing it to have a mean temperature of 2760\,K at $\sim$1.3\,R$_{\star}$, which is below both the photospheric effective temperature ($T_{\textrm{eff}} = 3690\,$K) and the temperatures at $\sim$$2\,$R$_{\star}$. This is unambiguous proof for the existence of an inversion of the mean temperature in the atmosphere of a red supergiant. The emission is clearly not spherically symmetric with two notable deviations from a uniform disk detected in both the images and visibilities. The most prominent asymmetry is located in the north-east quadrant of the disk and is spatially resolved showing it to be highly elongated with an axis-ratio of 2.4 and occupying $\sim$$5\%$ of the disk projected area. Its temperature is approximately 1000\,K above the measured mean temperature at 1.3\,R$_{\star}$. The other main asymmetry is located on the disk limb almost due east of the disk center and occupies $\sim$$3\%$ of the disk projected area. Both emission asymmetries are clear evidence for localized heating taking place in the atmosphere of Betelgeuse. We suggest that the detected localized heating is related to magnetic activity generated by large-scale photospheric convection.} 

   {}


   \maketitle
%
\section{Introduction}
The mechanisms responsible for driving the relatively large mass-loss rates ($\dot{M}\,\sim\,10^{-4}-10^{-6}\,M_{\odot}$\,yr$^{-1}$) of red supergiants (RSGs) remain poorly understood and the associated issues have long been highlighted \citep[e.g.,][]{holzer_1985}. RSGs have wind terminal velocities that are lower than their photospheric escape velocities (i.e., $v_{\infty} \lesssim 0.5v_{\textrm{esc}}$) and so most of the energy deposited into the wind does not go into the wind kinetic energy but must go into heating the atmosphere and lifting the atmosphere out of the gravitational potential well. An important constraint on the wind-driving mechanism of RSGs is that most of the energy must be deposited within the subsonic wind region to have any affect on mass-loss rates; energy deposited beyond merely modifies the wind velocity \citep{Hartmann_1980,holzer_1983}. Therefore, to gain insight into the mechanisms responsible for mass loss in RSGs, we need to study the atmosphere within the first few stellar radii (i.e., the extended atmosphere), where the wind gains most of its energy through heating and momentum deposition.  

Progress in this field is currently being driven by observations from state-of-art facilities that can provide the necessary spatial resolution to study the extended atmospheres of the closest RSGs. The Hubble Space Telescope (\textit{HST}) partially resolved the hot ($\sim$$8000\,$K) chromospheric gas around Betelgeuse ($\alpha$ Ori, M2Iab) and showed that the star in the ultraviolet (UV) appears about 2.5$\times$ larger than the optical photosphere (43\,mas in the infrared H-band, \citealt{montarges_2016}) and even more extended in the \ion{Mg}{II} H and K lines \citep{gilliland_1996}. The signature of an unresolved UV-bright asymmetry was also found and was suggested to be associated with the pole of the star \citep{uitenbroek_1998}. \cite{kervella_2016} imaged inhomogeneous hot H$\alpha$ emission mostly contained within 3\,R$_{\star}$ with SPHERE on the Very Large Telescope (VLT) while \cite{harper_2006} showed that the chromospheric gas decouples from co-rotation at about 3\,R$_{\star}$. \cite{lim_1998} used multi-wavelength, spatially resolved Very Large Array (VLA) radio observations to show that the hot chromospheric gas has a small filling factor and that the mean temperature at these radii is relatively cool ($T_{\textrm{gas}} \lesssim 3600\,$K). Indeed, the presence of a dense, cool (1500-2000\,K) molecular layer (so-called ``MOLsphere") is thought to exist between $\sim$1.3 and 1.5\,R$_{\star}$ to explain the infrared spectra of early M supergiants \citep[e.g.,][]{tsuji_2000,ohnaka_2011}. It appears that the extended atmospheres of RSGs exhibit complicated thermal structures.

For late-M RSGs like VY CMa, the dominant source of opacity at sub-mm  wavelengths is from dust grains \citep{ogorman_2015b}. However, for the earlier M-type RSGs such as Betelgeuse  the dominant source of opacity at sub-mm wavelengths is from H$^{-}$ opacity; thermal free-free interactions between free electrons from photoionized metals and neutral hydrogen \citep{harper_2001}. At these wavelengths, the thermal source function is simply the Planck function in the Rayleigh-Jeans limit, which is linear with the gas (electron) temperature. The mean gas temperature can then be empirically estimated using the Eddington-Barbier relation if the emission is spatially resolved. If the thermal structure is inhomogeneous at small scales (i.e., much smaller than the restoring beam), the source function is an average of the temperatures; unlike the UV, where very small volumes of hot plasma can dominate the total emission.

\section{Observations and analysis}
Betelgeuse was observed with the Atacama Large Millimeter/sub-millimeter Array (ALMA) on 9 November 2015 during Cycle 3 in Band 7 ($275 - 373\,$GHz) using an array configuration with projected baselines ranging from 0.078 to 16.076\,km  (project code: 2015.1.00206.S, PI: P. Kervella). The observations lasted for 75\,min with 61\,min spent on source. A total of 47 antennas were used to make the final calibrated visibility dataset. The observations were performed using five spectral windows (spws). Two spws were centered at 330.65\,GHz and 332.55\,GHz and covered a total bandwidth of 0.94\,GHz and 1.88\,GHz, respectively. The three other spws were centered at 343.19\,GHz, 345.15\,GHz, and 345.80\,GHz covering a total bandwidth of 1.88\,GHz, 0.94\,GHz and 0.47\,GHz, respectively.


The phase-reference source J0552+0313 was located 4.2$^{\circ}$ from Betelgeuse and was used to derive time-dependent phase and amplitude corrections. It was observed in 18\,s scans and bracketed 1.5\,min scans of Betelgeuse. The quasar J0510+1800 was used for bandpass and absolute flux density calibration assuming a flux density at 339.5\,GHz of 3.0\,Jy and a spectral index of $\alpha = -0.25$. We estimate an absolute flux density uncertainty of 5\% based upon the uncertainty of the catalog values due to  variability at the time. J0605+0939 was observed every 12 minutes and was used to check the absolute position of Betelgeuse (see \citealt{harper_2017}). The data were calibrated with the ALMA calibration pipeline and manually inspected and imaged using CASA 4.5.0. The continuum emission was split out from the molecular line emission and the continuum channel widths were averaged to 125\,MHz. A Gaussian profile was then fitted to the target in the initial continuum image to allow the target to be shifted to phase center. The total continuum bandwidth amounted to 4.1\,GHz at a center frequency of 338\,GHz ($\lambda = 0.89\,$mm). Continuum imaging is described in Appendix \ref{ap1}.  To analyze the self-calibrated visibilities directly, we used the UVMULTIFIT task version 2.2.0-r3 \citep{marti_vidal_2014}.
 

%

\begin{figure}[t]
\centering 
\includegraphics[trim=0pt 0pt 0pt 0pt,clip,angle=0, scale=0.46]{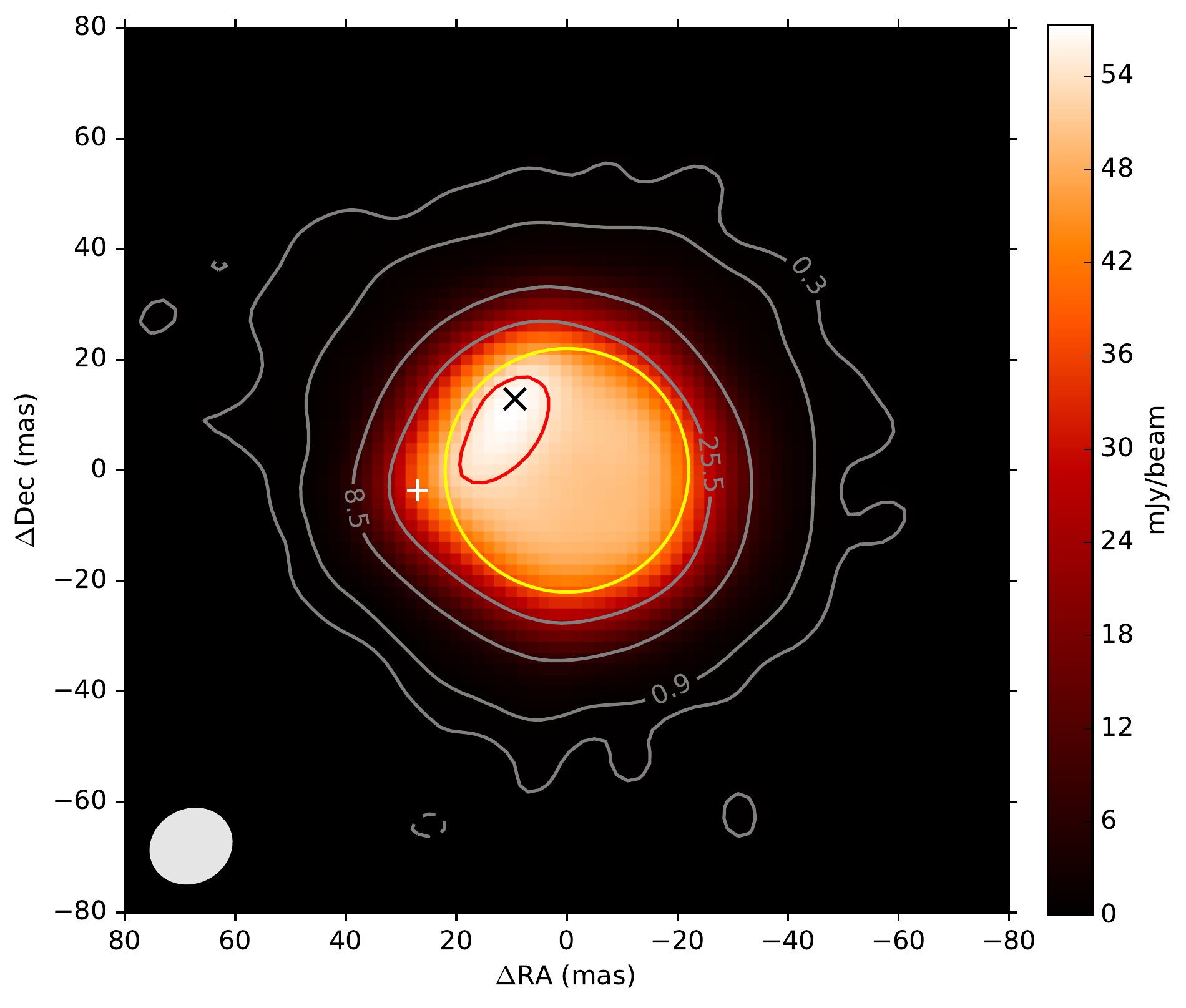}
\caption[]{ALMA 338\,GHz Briggs-weighted (robust 0) continuum image of Betelgeuse. Gray contours are set to $(-3, 3, 10, 100, 300) \times \sigma _{\textrm{rms}}$ where $\sigma _{\textrm{rms}} = 85\,\mu$Jy\,beam$^{-1}$ is the image rms noise. The numbers within the contours are the flux density values they represent in mJy. The largest contour is displayed in red for clarity at 54\,mJy or $635\sigma _{\textrm{rms}}$. The filled ellipse in the lower left corner is the full width at half maximum (FWHM) of the restoring beam and has dimensions $15\, \times \,13\,$mas. The locations of the NE-peak and the east-extension are marked with `$\times$' and `+', respectively. The yellow circle is the size of the infrared H-band photosphere from \cite{montarges_2016}.}
\label{fig1}
\end{figure}

\section{Results}
\begin{table*}
\caption{Best-fit multi-component model to the ALMA Band 7 continuum visibilities.}
\label{tab1}
\centering
\begin{tabular}{c c c c c c c c c}
\hline\hline
\rule{-3pt}{2ex}Component	& RA offset			& Dec. offset    & $S_{\nu}$           & Major axis       	& Axis ratio & PA & $T_{\textrm{b}}$	  \\
				&   (mas)			& (mas)	& 	 (mJy)        &   (mas)  & & ($^{\circ}$) & (K) \\
\hline
\rule{0pt}{2ex}Uniform elliptical disk & 0 & 0& 544.7$\pm$0.5 & 57.8$\pm$0.1 & 1.09$\pm$0.01 & 53.5$\pm$0.4 & 2760$\pm$140  \\
Elliptical Gaussian (NE-peak) & 9.4$\pm$0.1  & 12.9$\pm$0.1 & 16.1$\pm$0.4 & 19.8$\pm$0.5 &2.38$\pm$0.02 & 159.3$\pm$1.1 & 1055$\pm$90 \\
Circular Gaussian (east-extension)& 27.0$\pm$0.1 &-3.6$\pm$0.1 & 16.6$\pm$0.3 & 8.9$\pm$0.2 & 1& $\dots$& 2260$\pm$135 \\
Thin elliptical ring & 0 & 0 & 33.8$\pm$0.4 & 70.0$\pm$0.2 & 1.09$\pm$0.01 & 53.5$\pm$0.4 & $\dots$  \\
\hline
\end{tabular}
      \vspace{-2mm}
     \tablefoot{Positive right ascension (RA) and declination (Dec.) values represent east and north, respectively. $S_{\nu}$ is the flux density. Axis ratio = major axis/minor axis. Position angle (PA) is measured east of north. $T_{\textrm{b}}$ is the brightness temperature. Error values are the fitting errors except for the brightness temperature values which include the uncertainty in the absolute flux density.}
\end{table*}

The Briggs-weighted (robust 0) continuum image is shown in Figure \ref{fig1}. The restoring beam FWHM of $15\,\times\,13\,$mas clearly resolves the continuum emission with approximately 50 restoring beams fitting within the $3\sigma$ contours. The angular diameter of the emission at the $3\sigma$ level is approximately 100\,mas. The emission is clearly not axisymmetric with the peak emission notably located in the north-east (NE) quadrant (hereafter NE-peak). Moreover, the emission around the location of NE-peak itself is elongated. The more extended emission also deviates from circular symmetry with another notable asymmetry almost directly east of the center of emission (hereafter east-extension). The slightly higher spatial resolution images made using Briggs (robust -1) and uniform weighting yield almost an identical morphology as in Figure \ref{fig1}, therefore we exclude them from any further analysis and discussion due to their lower S/N.

We fitted a number of models to the complex visibilities to accurately quantify the basic properties of the main emission features such as position, morphology, and flux density, and also search for any underlying concealed substructure that might not be obvious in the images. The models included a uniform elliptical disk either with or without a combination of point sources, elliptical or circular Gaussians, and an elliptical ring. After each fit, the residual images (i.e., data - model) were inspected to help guide the next possible improved fit. Some of these residual images are shown in Appendix \ref{ap0} along with further details of the fitting process. The model that produced the lowest reduced chi-squared statistic value as well as the lowest remaining residuals in the residual images is described in Table \ref{tab1}. It consists of one relatively large uniform elliptical disk that contains one resolved elliptical Gaussian at the position of the NE-peak, another resolved circular Gaussian at the position of the east-extension, and a thin (i.e., unresolved) elliptical ring outside of the uniform elliptical disk with the same center position and position angle (PA; measured east of north) as the main uniform elliptical disk. 

The main uniform elliptical disk has a major axis of 57.8\,mas, is mildly elongated with an axis ratio (major axis/minor axis) of 1.09 and has a brightness temperature, $T_{\textrm{b}}$, of $2760\pm140$\,K. The NE-peak was found to be highly elongated having a FWHM major axis of 19.8\,mas and an axis ratio of 2.38. This elongation is clearly visible in both the continuum images and in the residual images where the component has not been subtracted (see the left panel of Figure \ref{fig2}). The total brightness temperature of the NE-peak was $3815\pm165$\,K. This value is derived by adding its $T_\textrm{b}$ value in Table \ref{tab1} to the uniform elliptical disk value because the model fitting has shown the entire NE-peak lies directly on top of the uniform elliptical disk. The east-extension was characterized by a circular Gaussian profile centered almost exactly at the limb/edge of the main uniform elliptical disk. Its FWHM was 8.9\,mas and its brightness temperature was $2260\pm 135$\,K. The total brightness temperature of the east-extension in reality will be higher than this value because a portion of it lies on top of the uniform elliptical disk. We attribute the thin ring of emission beyond the uniform intensity disk as emission that is not accounted for in the idealized uniform disk model. The ring is not located close to the limb of the main disk which suggests that it is not a result of limb brightening. The lack of a detection of limb brightening means that the thermal gradients are small on the scale height being probed when the finite resolution is accounted for.

The continuum data have only $\sim$$4\%$ fractional bandwidth coverage but the high S/N allows the spectral index, $\alpha$ (i.e., $S_{\nu} \propto \nu ^{\alpha}$), of the emission to be constrained. We find the global spectral index to be $\alpha = 1.67\,\pm 0.09 $ by separately imaging the lower (at 331.86\,GHz) and upper (at 344.09\,GHz) portions of the bandpass and calculating the integrated flux density in each image. This is slightly larger than the $\alpha = 1.57 \pm 0.05$ that can be derived from previous unresolved millimeter observations \citep{altenhoff_1994,harper_2009,ogorman_2012} and is much larger than the $\alpha = 1.33 \pm 0.01$ that can be derived from multi-epoch centimeter studies \citep{ogorman_2015a}. We note that the NE-peak has a spectral index of $\lesssim 2$ in the spectral index image (from CASA's clean parameter nterms = 2) but the error values per pixel may be underestimated due to the small fractional bandwidth coverage. Both the global and pixel spectral indices are in agreement with the emission being thermal free-free in nature, while the increasing $\alpha$ towards higher frequencies is a manifestation of probing smaller density scale heights at higher frequencies. 

\section{Discussion and conclusion}
\subsection{Verification of a temperature inversion between the photosphere and chromosphere}
\begin{figure}[hbt!]
\centering 
\includegraphics[trim=0pt 0pt 0pt 27pt,clip,angle=90, scale=0.37]{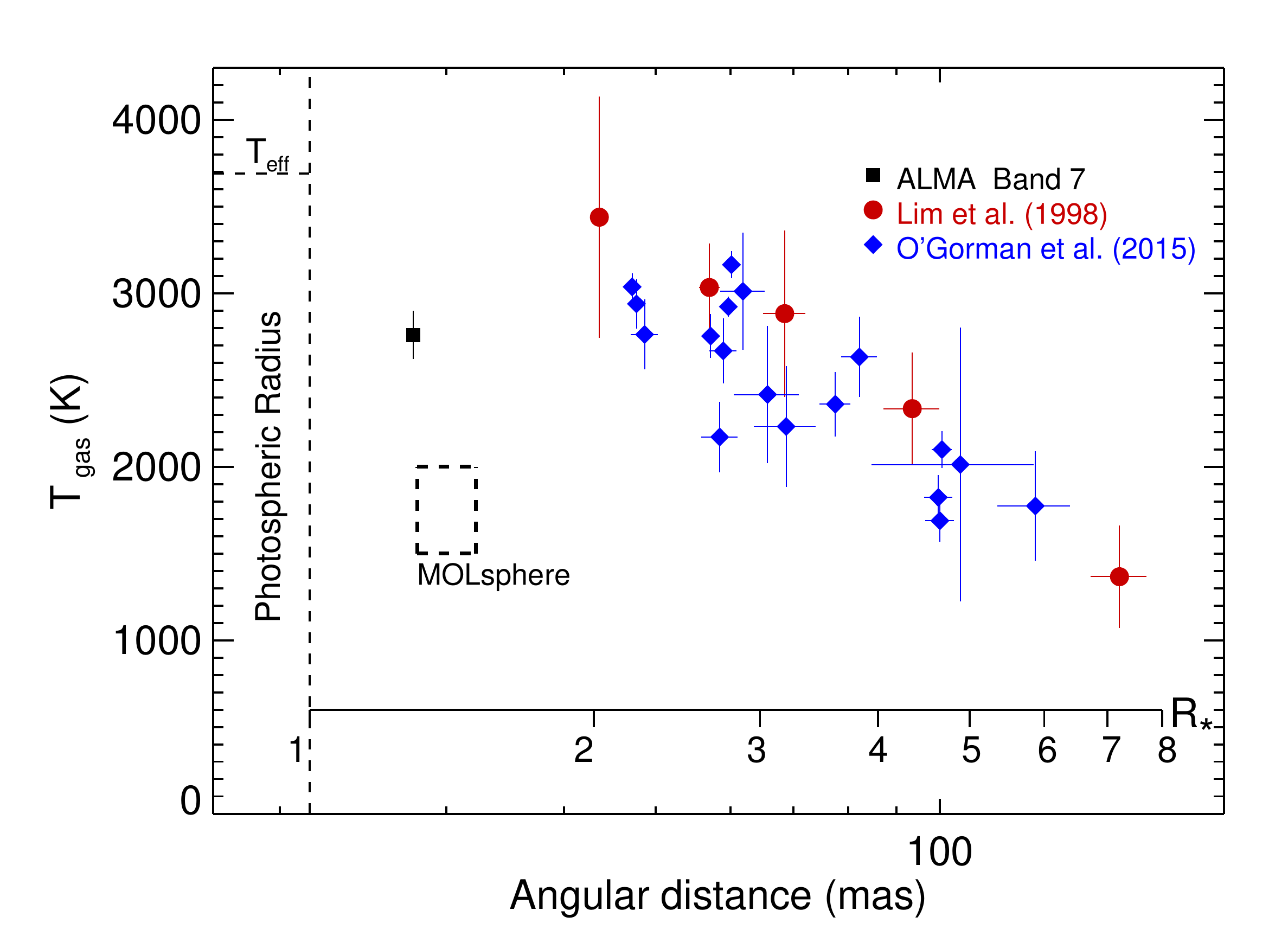}
\caption[]{Semi-log plot of the temperature profile of Betelgeuse's extended atmosphere. The red filled circles and blue filled diamonds represent the gas temperature derived from multi-epoch spatially resolved radio observations \citep{lim_1998,ogorman_2015a}. The large black dashed rectangle is the approximate location and temperature range of the MOLsphere. The black filled square is our ALMA band 7 temperature which shows that the mean gas temperature has dropped well below the effective temperature at $\sim$$1.3\,$R$_{\star}$.}
\label{fig2a}
\end{figure}

Sub-mm spatially resolved observations of thermal continuum emission from stellar atmospheres can act as an approximate linear thermometer. The continuum flux density, $S_{\nu}$, can be described as arising from an optically thick disk of angular size $\theta _{\nu}$ at some frequency $\nu$, and gas temperature $T_{\textrm{gas}}$ such that $S_{\nu} \propto T_{\textrm{gas}}\theta _{\nu} ^2$. At these frequencies the free-free opacity varies as roughly $\nu^2$, so multi-frequency observations allow the temperature profile to be constructed. This method has been used by \cite{lim_1998} and \cite{ogorman_2015a} to show that the mean gas temperature of Betelgeuse's extended atmosphere declines from approximately $3600\,$K at 2\,R$_{\star}$ to 1400\,K at 6\,R$_{\star}$, and is not as hot ($\sim$$8000\,$K) as previously thought \citep[e.g.,][]{hartmann_1984}. These values are plotted in Figure \ref{fig2a} along with our ALMA measurement of $2760\pm140\,$K at $\sim$$1.3\,$R$_{\star}$. This value is below the photospheric effective temperature of $3690\,\pm\,54\,$K \citep{ohnaka_2011} and most of the temperature measurements between $\sim$2 and 3\,R$_{\star}$. Obviously, there is not a monotonic decrease in the mean gas temperature of the extended atmosphere from the photosphere outwards. Indeed, modeling of infrared molecular lines from RSGs has already provided evidence for the presence of cool ($1500-2000\,$K) gas between 1.3 and 1.5\,R$_{\star}$ \citep{tsuji_2000, perrin_2004, perrin_2007, ohnaka_2004, montarges_2014}. However, these models are not sensitive to hotter plasma that may be co-located with the cool gas. Our ALMA measurement provides the mean gas temperature and is the reason why our value at $\sim$$1.3\,$R$_{\star}$ is larger in comparison to MOLsphere values.

\cite{lim_1998} suggested that photospheric-like temperatures at \textit{small} stellar radii (i.e., at $\sim$2\,R$_{\star}$), along with the monotonic decrease in temperature with increasing distance from the star, could be explained by the expansion and cooling of material elevated from the photosphere by large convection cells. In this scenario, the mean gas temperature should not drop below the effective  temperature between 1 and 2\,R$_{\star}$, but we find that it does. Our finding is in agreement with chromospheric modeling of all cool stars, which suggests that there is a trend of decreasing gas temperature above the photosphere which then rises again into the chromosphere \cite[e.g.,][]{basri_1981}. 

\subsection{The source of the sub-mm continuum asymmetries}
At the depths probed by the 338~GHz emission the gas temperature will vary much more slowly with height above the surface, $z$, than the hydrogen $(n_\textrm{H}$) and electron ($n_\textrm{e}$) densities, respectively \citep{basri_1981}. Eddington-Barbier relations show that for temperatures varying slowly with optical depth $T_\textrm{b}\simeq T_\textrm{gas}(\tau=1)$, where
\begin{equation}
\tau\left(z\right) \propto \int\limits_{z}^{\infty} n_\textrm{e} n_\textrm{H} \>dz^\prime
\simeq n_\textrm{e}^2(z)\> H_\rho/2 
\label{eq1}
\end{equation}
where $H_\rho$ is the density scale height. The latter equality follows if both $n_\textrm{e}/n_\textrm{H}$ and $H_\rho$ are approximately constant. This demonstrates that the sub-mm opacity is dominated by the local densities {\em and} the local scale-height. Localized (i.e., non-uniform) heating will not only increase the gas temperature but it will also increase $n_\textrm{e}$ (from photoionized metals), the density scale-height, and thus the total optical depth. This could explain why the NE-peak is the brightest feature in our image of Betelgeuse. On the stellar disk we are sampling $T_{\textrm{gas}}= 2760\,$K, but if the temperature increases outwards as a result of local heating, then the $\tau=1$ surface moves to a greater height where $T_{\textrm{gas}} = 3815$\,K. The east-extension can then be explained by the same phenomenon but only observed on the limb.

\cite{uitenbroek_1998} tentatively associated a strong UV hot spot with the location of the stellar rotation pole. \cite{uitenbroek_1998} and \cite{dupree_2011} suggested that global shock waves generated by pulsations would be more intense at the pole because of the steeper density gradients caused by stellar rotation. They suggested this scenario as the heating mechanism behind their detected UV hot spots. However, our detected regions of localized heating do not coincide with the UV hot spot, making the physical association with the pole less likely.

Solar sub-mm observations probe the lower chromosphere immediately above the temperature minimum. Applying the opacity scaling laws from \cite{ayres_1979} to Betelgeuse shows that its chromosphere has a lower opacity than the Sun and so our ALMA observations are most likely probing very close to the temperature minimum. Solar sub-mm studies show brightness temperature fluctuations across the solar disk with values $\sim$$1000\,$K in excess of the quiet Sun brightness in regions of strong magnetic field strength \citep{lindsey_1995,lindsey_1995a}. Given that Betelgeuse has a longitudinal magnetic field strength of $\sim1$G \citep{auriere_2010}, which is similar to the solar value, and it is predicted to harbor local regions of strong magnetic fields \citep{dorch_2004}, it is possible that the sub-mm asymmetries are related to magnetic activity. Near-infrared interferometric observations \citep[e.g.,][]{montarges_2016, houbois_2009} suggest there are large convection cells on the surface of Betelgeuse. \cite{dorch_2004} carried out 3D convection simulations for Betelgeuse and showed that these cells could sustain a local magnetic field. Indeed, a bright spot linked to a large convection cell has been reported in the north-east quadrant of Betelgeuse's photosphere less than one month prior to our ALMA observations \citep{tessore_2017}, which is tantalizing evidence for a link between photospheric convection and heating in the atmosphere of a RSG.

Our suggestion that the detected localized heating is due to magnetic activity would be in agreement with the conclusion in \cite{harper_2006} that very thin `filamentary' structures are required so that the chromospheric component is not completely opaque and detectable at cm wavelengths. These filaments would have a very small filling factor but could have gas temperatures similar to those implied by UV chromospheric continuum and lines studies (i.e., $T_{\textrm{gas}}$$\sim$$8000\,$K). However, our limited spatial resolution would smear out these fine structures and would reduce the contrast in the brightness temperatures. Indeed, \cite{loukitcheva_2015} have shown that this will happen when active regions on the Sun are observed at low spatial resolution with ALMA. Future multi-frequency ALMA observations with similar spatial resolution to that presented here could confirm our hypothesis if the brightness temperatures of the asymmetries increase with decreasing frequency as a result of the shifting of contributing heights to higher and hotter layers in the atmosphere.

\begin{acknowledgements}
This paper makes use of the following ALMA data: ADS/JAO.ALMA\#2015.1.00206.S. ALMA is a partnership of ESO (representing its member states), NSF (USA) and NINS (Japan), together with NRC (Canada), NSC and ASIAA (Taiwan), and KASI (Republic of Korea), in cooperation with the Republic of Chile. The Joint ALMA Observatory is operated by ESO, AUI/NRAO and NAOJ. We thank the Nordic ALMA Node for providing computational resources for this project. EOG acknowledges support from the Irish Research Council. We acknowledge financial support from the ``Programme National de Physique Stellaire" (PNPS) of CNRS/INSU, France. GMH received support from HST grant  HST -AR-14566.001-A which was provided by NASA through a grant from the Space Telescope Science Institute, which is operated by the Association of Universities for Research Astronomy, Incorporated, under NASA contract NAS5-26555.
\end{acknowledgements}

\bibliographystyle{aa}
\bibliography{references}

\newcommand{\noop}[1]{}
\begin{thebibliography}{34}
\expandafter\ifx\csname natexlab\endcsname\relax\def\natexlab#1{#1}\fi

\bibitem[{{Altenhoff} {et~al.}(1994){Altenhoff}, {Thum}, \&
  {Wendker}}]{altenhoff_1994}
{Altenhoff}, W.~J., {Thum}, C., \& {Wendker}, H.~J. 1994, \aap, 281, 161

\bibitem[{{Auri{\`e}re} {et~al.}(2010){Auri{\`e}re}, {Donati},
  {Konstantinova-Antova}, {Perrin}, {Petit}, \& {Roudier}}]{auriere_2010}
{Auri{\`e}re}, M., {Donati}, J.-F., {Konstantinova-Antova}, R., {et~al.} 2010,
  \aap, 516, L2

\bibitem[{{Ayres}(1979)}]{ayres_1979}
{Ayres}, T.~R. 1979, \apj, 228, 509

\bibitem[{{Basri} {et~al.}(1981){Basri}, {Eriksson}, \& {Linsky}}]{basri_1981}
{Basri}, G.~S., {Eriksson}, K., \& {Linsky}, J.~L. 1981, \apj, 251, 162

\bibitem[{{Dorch}(2004)}]{dorch_2004}
{Dorch}, S.~B.~F. 2004, \aap, 423, 1101

\bibitem[{{Dupree}(2011)}]{dupree_2011}
{Dupree}, A.~K. 2011, in IAU Symposium, Vol. 273, Physics of Sun and Star
  Spots, ed. D.~{Prasad Choudhary} \& K.~G. {Strassmeier}, 188--194

\bibitem[{{Gilliland} \& {Dupree}(1996)}]{gilliland_1996}
{Gilliland}, R.~L. \& {Dupree}, A.~K. 1996, \apjl, 463, L29

\bibitem[{{Harper} \& {Brown}(2006)}]{harper_2006}
{Harper}, G.~M. \& {Brown}, A. 2006, \apj, 646, 1179

\bibitem[{{Harper} {et~al.}(\noop{3001}in press){Harper}, {Brown}, {Guinan},
  {O'Gorman}, {Richards}, {Kervella}, \& {Decin}}]{harper_2017}
{Harper}, G.~M., {Brown}, A., {Guinan}, E.~F., {et~al.} \noop{3001}in press,
  \aj

\bibitem[{{Harper} {et~al.}(2001){Harper}, {Brown}, \& {Lim}}]{harper_2001}
{Harper}, G.~M., {Brown}, A., \& {Lim}, J. 2001, \apj, 551, 1073

\bibitem[{{Harper} {et~al.}(2009){Harper}, {Richter}, {Ryde}, {Brown}, {Brown},
  {Greathouse}, \& {Strong}}]{harper_2009}
{Harper}, G.~M., {Richter}, M.~J., {Ryde}, N., {et~al.} 2009, \apj, 701, 1464

\bibitem[{{Hartmann} \& {Avrett}(1984)}]{hartmann_1984}
{Hartmann}, L. \& {Avrett}, E.~H. 1984, \apj, 284, 238

\bibitem[{{Hartmann} \& {MacGregor}(1980)}]{Hartmann_1980}
{Hartmann}, L. \& {MacGregor}, K.~B. 1980, \apj, 242, 260

\bibitem[{{Haubois} {et~al.}(2009){Haubois}, {Perrin}, {Lacour}, {Verhoelst},
  {Meimon}, {Mugnier}, {Thi{\'e}baut}, {Berger}, {Ridgway}, {Monnier},
  {Millan-Gabet}, \& {Traub}}]{houbois_2009}
{Haubois}, X., {Perrin}, G., {Lacour}, S., {et~al.} 2009, \aap, 508, 923

\bibitem[{{Holzer} {et~al.}(1983){Holzer}, {Fla}, \& {Leer}}]{holzer_1983}
{Holzer}, T.~E., {Fla}, T., \& {Leer}, E. 1983, \apj, 275, 808

\bibitem[{{Holzer} \& {MacGregor}(1985)}]{holzer_1985}
{Holzer}, T.~E. \& {MacGregor}, K.~B. 1985, in Astrophysics and Space Science
  Library, Vol. 117, Mass Loss from Red Giants, ed. M.~{Morris} \&
  B.~{Zuckerman}, 229--255

\bibitem[{{Kervella} {et~al.}(2016){Kervella}, {Lagadec}, {Montarg{\`e}s},
  {Ridgway}, {Chiavassa}, {Haubois}, {Schmid}, {Langlois}, {Gallenne}, \&
  {Perrin}}]{kervella_2016}
{Kervella}, P., {Lagadec}, E., {Montarg{\`e}s}, M., {et~al.} 2016, \aap, 585,
  A28

\bibitem[{{Lim} {et~al.}(1998){Lim}, {Carilli}, {White}, {Beasley}, \&
  {Marson}}]{lim_1998}
{Lim}, J., {Carilli}, C.~L., {White}, S.~M., {Beasley}, A.~J., \& {Marson},
  R.~G. 1998, \nat, 392, 575

\bibitem[{{Lindsey} \& {Kopp}(1995)}]{lindsey_1995a}
{Lindsey}, C. \& {Kopp}, G. 1995, \apj, 453, 517

\bibitem[{{Lindsey} {et~al.}(1995){Lindsey}, {Kopp}, {Clark}, \&
  {Watt}}]{lindsey_1995}
{Lindsey}, C., {Kopp}, G., {Clark}, T.~A., \& {Watt}, G. 1995, \apj, 453, 511

\bibitem[{{Loukitcheva} {et~al.}(2015){Loukitcheva}, {Solanki}, {Carlsson}, \&
  {White}}]{loukitcheva_2015}
{Loukitcheva}, M., {Solanki}, S.~K., {Carlsson}, M., \& {White}, S.~M. 2015,
  \aap, 575, A15

\bibitem[{{Mart{\'{\i}}-Vidal} {et~al.}(2014){Mart{\'{\i}}-Vidal}, {Vlemmings},
  {Muller}, \& {Casey}}]{marti_vidal_2014}
{Mart{\'{\i}}-Vidal}, I., {Vlemmings}, W.~H.~T., {Muller}, S., \& {Casey}, S.
  2014, \aap, 563, A136

\bibitem[{{Montarg{\`e}s} {et~al.}(2016){Montarg{\`e}s}, {Kervella}, {Perrin},
  {Chiavassa}, {Le Bouquin}, {Auri{\`e}re}, {L{\'o}pez Ariste}, {Mathias},
  {Ridgway}, {Lacour}, {Haubois}, \& {Berger}}]{montarges_2016}
{Montarg{\`e}s}, M., {Kervella}, P., {Perrin}, G., {et~al.} 2016, \aap, 588,
  A130

\bibitem[{{Montarg{\`e}s} {et~al.}(2014){Montarg{\`e}s}, {Kervella}, {Perrin},
  {Ohnaka}, {Chiavassa}, {Ridgway}, \& {Lacour}}]{montarges_2014}
{Montarg{\`e}s}, M., {Kervella}, P., {Perrin}, G., {et~al.} 2014, \aap, 572,
  A17

\bibitem[{{O'Gorman} {et~al.}(2015{\natexlab{a}}){O'Gorman}, {Harper}, {Brown},
  {Guinan}, {Richards}, {Vlemmings}, \& {Wasatonic}}]{ogorman_2015a}
{O'Gorman}, E., {Harper}, G.~M., {Brown}, A., {et~al.} 2015{\natexlab{a}},
  \aap, 580, A101

\bibitem[{{O'Gorman} {et~al.}(2012){O'Gorman}, {Harper}, {Brown}, {Brown},
  {Redfield}, {Richter}, \& {Requena-Torres}}]{ogorman_2012}
{O'Gorman}, E., {Harper}, G.~M., {Brown}, J.~M., {et~al.} 2012, \aj, 144, 36

\bibitem[{{O'Gorman} {et~al.}(2015{\natexlab{b}}){O'Gorman}, {Vlemmings},
  {Richards}, {Baudry}, {De Beck}, {Decin}, {Harper}, {Humphreys}, {Kervella},
  {Khouri}, \& {Muller}}]{ogorman_2015b}
{O'Gorman}, E., {Vlemmings}, W., {Richards}, A.~M.~S., {et~al.}
  2015{\natexlab{b}}, \aap, 573, L1

\bibitem[{{Ohnaka}(2004)}]{ohnaka_2004}
{Ohnaka}, K. 2004, \aap, 421, 1149

\bibitem[{{Ohnaka} {et~al.}(2011){Ohnaka}, {Weigelt}, {Millour}, {Hofmann},
  {Driebe}, {Schertl}, {Chelli}, {Massi}, {Petrov}, \& {Stee}}]{ohnaka_2011}
{Ohnaka}, K., {Weigelt}, G., {Millour}, F., {et~al.} 2011, \aap, 529, A163

\bibitem[{{Perrin} {et~al.}(2004){Perrin}, {Ridgway}, {Coud{\'e} du Foresto},
  {Mennesson}, {Traub}, \& {Lacasse}}]{perrin_2004}
{Perrin}, G., {Ridgway}, S.~T., {Coud{\'e} du Foresto}, V., {et~al.} 2004,
  \aap, 418, 675

\bibitem[{{Perrin} {et~al.}(2007){Perrin}, {Verhoelst}, {Ridgway}, {Cami},
  {Nguyen}, {Chesneau}, {Lopez}, {Leinert}, \& {Richichi}}]{perrin_2007}
{Perrin}, G., {Verhoelst}, T., {Ridgway}, S.~T., {et~al.} 2007, \aap, 474, 599

\bibitem[{{Tessore} {et~al.}(2017){Tessore}, {L{\`o}pez-Ariste}, {Mathias},
  {L{\`e}bre}, {Morin}, \& {Josselin}}]{tessore_2017}
{Tessore}, B., {L{\`o}pez-Ariste}, A., {Mathias}, P., {et~al.} 2017, ArXiv
  e-prints [\eprint[arXiv]{1702.02002}]

\bibitem[{{Tsuji}(2000)}]{tsuji_2000}
{Tsuji}, T. 2000, \apj, 538, 801

\bibitem[{{Uitenbroek} {et~al.}(1998){Uitenbroek}, {Dupree}, \&
  {Gilliland}}]{uitenbroek_1998}
{Uitenbroek}, H., {Dupree}, A.~K., \& {Gilliland}, R.~L. 1998, \aj, 116, 2501

\end{thebibliography}

\begin{appendix} 

\section{Continuum imaging}\label{ap1}
Continuum imaging was performed using the multi-frequency synthesis CLEAN algorithm with Briggs weighting and a robust parameter of 0, which resulted in a synthetic beam size of $15\,\times \,13\,$mas at 338\,GHz. Two Taylor coefficients were used to model the linear frequency dependence of the continuum emission. Additionally, we used the multiscale imaging option with scales approximately corresponding to 0, 1, and $3\,\times$ the synthesized beam. Using these imaging parameters, three iterations of phase-only self-calibration were performed on the continuum emission until a S/N convergence was reached with a solution interval of 15\,sec. One final iteration was then performed, solving for both amplitude and phase. We note that self-calibrating the continuum data reduced the rms noise by approximately an order of magnitude. To investigate small-scale structure in the continuum emission we also created images using Briggs weighting with a robust parameter of -1 and uniform weighting. The achieved rms noise for the continuum images was 85\,$\mu$Jy\,beam$^{-1}$ (robust 0 images), 138\,$\mu$Jy\,beam$^{-1}$ (robust -1 images), and 255\,$\mu$Jy\,beam$^{-1}$ (uniform images).

\section{Residual images from \textit{uv}-fitting}\label{ap0}

\begin{figure*}[t!]
\centering
\includegraphics[trim=0pt 0pt 0pt 0pt,clip, scale=0.47]{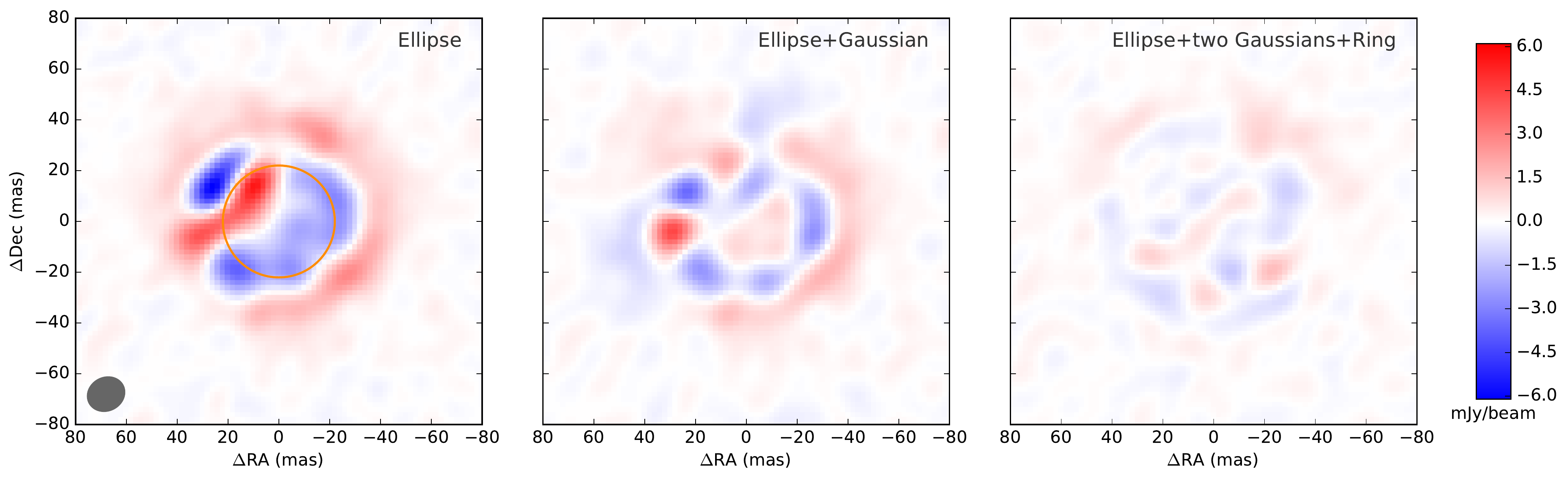}
\caption[]{ALMA 338\,GHz continuum residual (i.e., data - model) images of Betelgeuse. The filled ellipse in the lower left corner of the left panel is the FWHM of the restoring beam and has dimensions $15\, \times \,13\,$mas while the orange circle is the size of the infrared H-band photosphere from \cite{montarges_2016}. \textit{Left panel:} The residual image after a uniform elliptical disk has been subtracted from the visibilities. \textit{Middle panel:} The residual image after a uniform elliptical disk plus an elliptical Gaussian have been subtracted from the visibilities. \textit{Right panel:} The residual image after a uniform elliptical disk, an elliptical Gaussian, a circular Gaussian, and a thin ring have been subtracted from the visibilities.}
\label{fig2}
\end{figure*}

In Figure \ref{fig2} we show three example residual images from our \textit{uv}-fitting analysis. Residual images containing a lot of emission meant that the subtracted model did not describe the data well. In the left panel of Figure \ref{fig2} a uniform intensity elliptical disk was fitted to the visibilities. The best fit model was then subtracted from them using CASA's \textit{uvsub} task, and the residual visibilities were then imaged. The rms of the residual image within a circle of radius $\sim$3R$_{\star}$ centered on the disk center was 1.345\,mJy\,beam$^{-1}$. Clearly, excess emission remains in the residual image: A strong ($\sim$$\pm$60$\sigma _{\textrm{rms}}$) elongated feature in the north-east quadrant, another strong ($\sim$$\pm$40$\sigma _{\textrm{rms}}$) feature east of the disk center, and a weaker ($\sim$$\pm$10$\sigma$) ring-like feature beyond these. In the middle panel of Figure \ref{fig2} we show the residual image after the best fit model of a uniform intensity elliptical disk plus a Gaussian profile have been fitted and subtracted from the visibilities. The emission feature in the north-east quadrant is now gone and the rms of the residual image within a circle of radius $\sim$3R$_{\star}$ centered on the disk center was 0.817\,mJy\,beam$^{-1}$. Finally, in the right panel of Figure \ref{fig2} we show the residual image after the best fit model of a uniform intensity elliptical disk plus two Gaussian profiles plus a thin ring have been fitted and subtracted from the visibilities. All the main emission features present in the left panel of Figure \ref{fig2} are now gone and the rms of the residual image within a circle of radius $\sim$3R$_{\star}$ centered on the disk center was 0.367\,mJy\,beam$^{-1}$. We note that a number of small-scale features with ($\sim$$\pm5-10\sigma _{\textrm{rms}}$) significance remain, which are probably the result of the thin ring  not being a perfect match for the low brightness emission beyond the main elliptical disk.

We found that replacing the elliptical uniform intensity disk with a circular uniform intensity disk in the multi-component fits described previously produced larger chi-squared statistic values and increased the residual image rms. We also attempted fits by forcing the Gaussian components to have negative flux density values, but the fitted values were smaller than the fitting errors. Moreover, the large negative features in the left and middle panel are not present in our best fit model shown in the right panel and we conclude that they are simply artefacts.

\end{appendix}
\end{document}